\documentclass[conference]{IEEEtran}

\IEEEoverridecommandlockouts
\usepackage{cite}
\usepackage{amsmath,amssymb,amsfonts}
\usepackage{algorithmic}
\usepackage{graphicx}
\usepackage{textcomp}
\usepackage{xcolor}
\def\BibTeX{{\rm B\kern-.05em{\sc i\kern-.025em b}\kern-.08em
    T\kern-.1667em\lower.7ex\hbox{E}\kern-.125emX}}
\usepackage[colorlinks,urlcolor=blue,citecolor=blue]{hyperref}
\usepackage{multirow}
\usepackage[hyphenbreaks]{breakurl}

\makeatletter
\newcommand{\linebreakand}{%
  \end{@IEEEauthorhalign}
  \hfill\mbox{}\par
  \mbox{}\hfill\begin{@IEEEauthorhalign}
}
\makeatother

\begin{document}

\newcommand{\red}[1]{\textcolor{red}{[#1]}}

\title{Quantum Computing, Math, and Physics (QCaMP):\\Introducing quantum computing in high schools\\

\thanks{This material is based upon work supported by the U.S. Department of Energy, Office of Science, National Quantum Information Science Research Centers, Quantum Systems Accelerator. Additional support is acknowledged from Sandia National Laboratories, Lawrence Berkeley National Lab, and IEEE. SAND2023-09350O}
}

\author{\IEEEauthorblockN{Megan Ivory}
\IEEEauthorblockA{\textit{Photonic Microsystems Technology} \\
\textit{Sandia National Laboratories}\\
Albuquerque, NM, USA \\
mkivory@sandia.gov}
\and
\IEEEauthorblockN{Alisa Bettale}
\IEEEauthorblockA{\textit{Office of Government and} \\
\textit{Community Relations}\\
\textit{Lawrence Berkeley National Laboratory}\\
Berkeley, CA, USA \\
abettale@lbl.gov}
\and
\IEEEauthorblockN{Rachel Boren}
\IEEEauthorblockA{\textit{College of Health, Education,}\\
\textit{and Social Transformation} \\
\textit{New Mexico State University}\\
Las Cruces, NM, USA \\
rboren@nmsu.edu}
\linebreakand
\IEEEauthorblockN{Ashlyn D. Burch}
\IEEEauthorblockA{\textit{Photonic Microsystems Technology} \\
\textit{Sandia National Laboratories}\\
Albuquerque, NM, USA \\
adburch@sandia.gov}
\and
\IEEEauthorblockN{Jake Douglass}
\IEEEauthorblockA{\textit{Technical Business Development} \\
\textit{Sandia National Laboratories}\\
Albuquerque, NM, USA \\
jsdougl@sandia.gov}
\and
\IEEEauthorblockN{Lisa Hackett}
\IEEEauthorblockA{\textit{Photonic Microsystems Technology} \\
\textit{Sandia National Laboratories}\\
Albuquerque, NM, USA \\
lahacke@sandia.gov}
\linebreakand
\IEEEauthorblockN{Boris Kiefer}
\IEEEauthorblockA{\textit{Department of Physics} \\
\textit{New Mexico State University}\\
Las Cruces, NM, USA \\
bkiefer@nmsu.edu}
\and
\IEEEauthorblockN{Alina Kononov}
\IEEEauthorblockA{\textit{Quantum Algorithms and Applications} \\
\textit{Collaboratory} \\
\textit{Sandia National Laboratories}\\
Albuquerque, NM, USA \\
akonono@sandia.gov}
\and
\IEEEauthorblockN{Maryanne Long}
\IEEEauthorblockA{\textit{College of Health, Education,}\\
\textit{and Social Transformation} \\
\textit{New Mexico State University}\\
Las Cruces, NM, USA \\
long2@nmsu.edu}
\linebreakand
\IEEEauthorblockN{Mekena Metcalf}
\IEEEauthorblockA{\textit{Innovation and Ventures} \\
\textit{HSBC, Holdings Plc.}\\
San Francisco, CA, USA \\
mekena.metcalf@us.hsbc.com}
\and
\IEEEauthorblockN{Tzula B. Propp}
\IEEEauthorblockA{\textit{{Center for Quantum Information \& Control}} \\
\textit{University of New Mexico}\\
Albuquerque, NM, USA \\
tzpropp@unm.edu}
\and
\IEEEauthorblockN{Mohan Sarovar}
\IEEEauthorblockA{\textit{Quantum Algorithms and Applications}\\
\textit{Collaboratory} \\
\textit{Sandia National Laboratories}\\
Livermore, CA, USA \\
mnsarov@sandia.gov}
}

\maketitle

\begin{abstract}
The nascent but rapidly growing field of Quantum Information Science and Technology has led to an increased demand for skilled quantum workers and an opportunity to build a diverse workforce at the outset.  In order to meet this demand and encourage women and underrepresented minorities in STEM to consider a career in QIST, we have developed a curriculum for introducing quantum computing to teachers and students at the high school level with no prerequisites.  In 2022, this curriculum was delivered over the course of two one-week summer camps, one targeting teachers and another targeting students.  Here, we present an overview of the objectives, curriculum, and activities, as well as results from the formal evaluation of both camps and the outlook for expanding QCaMP in future years. 
\end{abstract}

\begin{IEEEkeywords}
Quantum Information Science and Technology, Quantum Education, Quantum Outreach
\end{IEEEkeywords}

\section{Introduction}
Quantum Information Science and Technology (QIST) is a nascent field that has been recognized by the US government as a key strategic investment area due to its promising applications in sensing, time-keeping, computing, and communications, and it is set to see significant investment in the years to come \cite{QISTWD2022,QEDC2022}. As the field expands, so do the demands for a workforce with the skills, expertise, and educational experiences needed to meet the ever-evolving requirements.  There is broad recognition that a critical need exists to grow the QIST workforce across the full educational spectrum – not just at the PhD level – and to engage new communities if the US is to maintain leadership in the field \cite{QISTWD2022}.

Current efforts to address the QIST workforce demand largely focus on graduate level programs and upskilling quantum-adjacent professionals from computer science, engineering, math, and other STEM-fields. Unfortunately these fields traditionally lack diversity \cite{DinSTEM2023}.  We believe that introducing QIST concepts at the high school level will enable us to reach a more diverse group of students, and this awareness will enable students to be better prepared for a career in QIST.

To that end, we launched the Quantum Computing, Math, and Physics camp (QCaMP) in 2022 to introduce QIST concepts to high school teachers and students with the following goals:

\begin{itemize}
    \item Introduce topics through meaningful hands-on puzzles, experiments, and demos with no advanced mathematics or prerequisites,
    \item Provide teachers with the tools, experience, and modules that are in line with classroom standards and easy to integrate into high school curricula, 
    \item Expose students to different career pathways, 
    \item Reach and relate to participants with diverse demographics, including gender, racial/ethnic, regional, and socio-economic diversity.
\end{itemize}

QCaMP brought together 27 contributors from 8 institutions spanning national laboratories, academia, industry, and educational non-profits to teach 32 students and 20 teachers about QIST over two one-week-long virtual camps. To encourage socio-economic diversity, materials were sent to participants free-of-charge and each participant received a stipend.  Financial stress disproportionately affects the underrepresented communities we hope to reach with this program \cite{finaid}.  As such, prospective participants in programs like this one face the decision of working summer jobs to support themselves and their families vs attending summer camps that could prepare them for better educational and career opportunities.  We sought to mitigate these financial impacts by providing teachers with $\$1000$ for the week and students with $\$250$ for the week.  Stipend amounts were chosen to be comparable to a 1-week teacher salary in NM and to half-week at minimum wage for teachers and students, respectively. These stipends made up the majority of QCaMP's costs.

We present an overview of our curriculum in Section \ref{sec:curriculum}, outcomes from the pilot camp evaluations in Section \ref{sec:outcomes}, and outlooks for further refining and expanding the program in Section \ref{sec:outlook}.

\section{Curriculum}
\label{sec:curriculum}

QIST researchers from national laboratories and academia collaboratively developed the curriculum around topics shown in Table \ref{Curr}.  Our aim was to make the material accessible to a broad range of students by avoiding any prerequisite math, physics, or computer science courses and to incorporate experiential and active learning wherever possible.  Beginning with no assumed background knowledge, the curriculum introduces students to relevant topics, culminating in a research project using IBM's cloud-based quantum computer (IBM Q composer) at the end of the week.

\begin{table*}[htbp]
\caption{QCaMP 2022 Curriculum}
\begin{center}
\begin{tabular}{|c|c|c|c|c|c|}
\hline
& \textbf{Monday} & \textbf{Tuesday} & \textbf{Wednesday} & \textbf{Thursday} & \textbf{Friday}\\
\hline
\multirow{2}{*}{\textbf{AM}} & Classical Bits and Gates & Double Slit Experiment & Superposition & Entanglement & IBM Q Research Project \\
& (Section \ref{sec:bits}) & (Section \ref{sec:lightlabs}) & (Section \ref{sec:superposition}) & (Section \ref{sec:entanglement}) & (Section \ref{sec:ibm}) \\
\hline
\multirow{2}{*}{\textbf{PM}} & Probability and Statistics & Polarization Experiment & Intro to IBM Q & IBM Q Research Project & Virtual Lab Tours \\
& (Section \ref{sec:probability}) & (Section \ref{sec:lightlabs}) & (Section \ref{sec:ibm}) & (Section \ref{sec:ibm}) & (Section \ref{sec:othersessions})\\
\hline
\end{tabular}
\label{Curr}
\end{center}
\end{table*}

The curricula for the teacher camp and student camp were largely the same. Key differences include the following: teacher camps included additional time for pedagogical discussions and resource sharing while student camps included career talks and follow-on extracurricular educational opportunities. These topics were covered during discussion and Q\&A time and via mini-talks at the end of each day. For both camps, each day ended with a Q\&A session utilizing Padlet.com. This tool offered an anonymous platform for soliciting feedback, questions, and comments from the participants. The following morning, we addressed the common misconceptions or areas that needed additional instruction.  
In the following, we describe each module in more detail.

\subsection{Classical Bits and Gates}
\label{sec:bits}

On Monday morning, we began by introducing classical computing fundamentals, including bits and gates.  We used the formalism for bits and gates developed in \cite{Qi4Q} and utilized in \cite{sophia} -- white and black balls representing 0 and 1 bits, respectively, which are dropped through various boxes representing gates.  Basic gates including NOT, CNOT, and SWAP were introduced and reinforced through guided exercises in which participants worked out truth tables for individual gates and determined the output of various combinations of gates.  We also discussed how information could be stored in anything with two mutually exclusive states, setting the stage for the later lab tours showcasing different qubit hardware types.
Finally, we acknowledged that current hardware remains noisy and described simple schemes for error detection and correction.

\subsection{Probability and Statistics}
\label{sec:probability}

We acknowledged that students would need some understanding of probability and statistics in order to interpret and understand randomness in the outcomes of quantum experiments. A very short and practical introduction to probability and statistics was provided on Monday afternoon. We found that although most students did not have formal training in these topics, they had intuition for the concept of probability and could reason about events, likelihoods, \emph{etc.} The probability portion of the lesson explained basic combinatorics principles, including mutually exclusive and independent events. Then, the statistics portion focused on practical computations the participants would perform in the hands-on portions of the camp, especially in the final project's analysis of quantum circuits run on IBM's cloud-based quantum processors. We introduced concepts such as mean, variance, and standard deviation; provided intuition for each of these quantities; and then demonstrated how they can be computed on real data sets using the Google Sheets spreadsheet application.

\subsection{Light Labs}
\label{sec:lightlabs}

On Tuesday, we included more physics-related topics using hands-on activities that had students and teachers step away from the computer. We provided participants with necessary materials prior to the start of the camp, shipping them lasers, apertures, and polarizers that they could use in their experiments (see Section \ref{sec:materials} for more details about these kits). 
We began by exploring how light passes through apertures. We also used the University of Colorado at Boulder's PhET Interactive Simulations \cite{phet} to hone in on these concepts which culminated in a double slit experiment by the end of the morning.  Using the concepts they learned performing these lab experiments, we connected the wave-particle nature of light to their observations and began to introduce the concepts of superposition and measurement. 

Tuesday afternoon was devoted to polarization in a lesson adapted from the University of Waterloo's Schrodinger's Class materials \cite{Waterloo}. In order to accommodate students who have not yet taken trigonometry, a more conceptual approach was taken, exploring phenomena through direct observations of the effects of polarizers. The activity started by exploring how a single polarizer blocks light from a computer screen and any other sources of light that teachers and students wanted to observe. To connect back to Monday's lessons, the topic of mutually exclusive states was reintroduced, highlighting the fact that polarization provides mutually exclusive states that can store information. 
Next, teachers and students used two polarizers to directly observe which polarization states were mutually exclusive, followed by an introduction of the concept of measurement to make sense of their observations. Finally, the camp participants observed the surprising effect of inserting a diagonal polarizer between vertical and horizontal polarizers, bringing in the concept of superposition and relating it again to measurement.

\subsection{Superposition and the Hadamard Gate}
\label{sec:superposition}

On Wednesday morning, we formally introduced the concepts of superposition and measurement, drawing from examples camp participants encountered the day before in the lab session. We presented the differences between classical and quantum superposition and found that the most challenging portion of this lesson for both students and teachers was grasping the physical concept behind mixed quantum states. There was ample time dedicated at the beginning of this session for brainstorming and discussion on its physical realization in the world. Teachers were additionally provided content on how superposition could be represented using vectors, building upon high school trigonometric concepts of the unit circle to guide them to a basic understanding of a mixed state as it would be depicted in a Bloch sphere. Conceptual details about the Bloch sphere were presented to the students in a separate mini-talk (see Section \ref{sec:othersessions}), but teachers found the math very useful in this session. 

We then finally introduced the quantum Hadamard gate, building upon the foundation introduced on the first day with classical gates. In a similar vein, black and white colored balls represented observed qubits. Dropping them through the H box (i.e. Hadamard gate) produced a `misty' state as described in \cite{Qi4Q} where the balls emerged in some superposition of black and white. We included additional exercises combining classical and quantum gates to cement this learning. While performing these puzzles, the participants were also taught about the intrusive nature of measurement and the induced collapse of the `misty' state into one of the two observed black or white balls. 

\subsection{Entanglement}
\label{sec:entanglement}

Entanglement was introduced as a final module on Thursday morning.  While we initially developed an activity to measure the CHSH value using the QuVis simulations \cite{QuVis}, we realized during the teacher camp that this was much too advanced for our audience.  For the student camp, we significantly modified this lesson to conceptually discuss the differences between entanglement and classical correlation.  We also acknowledged that scientists are still debating the explanations behind these concepts as an effort to encourage similar discussion amongst the students.  
The many-worlds theory seemed to inspire the most discussion, as students could relate somewhat easily due to similar interpretations appearing in pop culture entertainment.

\subsection{IBM Q}
\label{sec:ibm}

The camp was structured to culminate in a final project using IBM Q, reflecting our guiding principle that the best way to learn how a quantum computer works is to actually use one. In fact, access to quantum hardware immediately distinguishes this field from its predecessor, classical computing. The invention of the Personal Computer (PC) came decades after computers became useful for solving complex problems. Computing access and network access for everyone revolutionized our society. More and more quantum computing architectures are becoming publicly available through the cloud and – typically – accessible to everyone. 

On Wednesday afternoon, we introduced camp participants to the IBM Quantum Composer \cite{IBM}, a visual drag-and-drop user interface that requires no prerequisite programming experience.  Teachers and students were familiarized with this tool first by reproducing various classical gates puzzles and using the individual qubit ``Prob of $|1\rangle$" readout as well as the State Vector plots to verify their results.  Later, these features were used to reinforce the results of puzzles incorporating the Hadamard gate.  
Finally, teachers and students learned how to perform measurements and run small single and multi-qubit circuits on IBM's publicly available quantum computers.
Building upon the probability and statistics lesson, they interpreted resulting histograms to understand read-out errors and CNOT errors.
Many participants were shocked to learn that real quantum computers existed today and could be accessed via the cloud for free.
 
 As a final project, on Thursday and Friday, we designed a hands-on research exercise using IBM Q that built upon the material learned throughout the camp, making quantum computing come to life and providing a practical and relevant tool for the students and teachers to use after the course. For this exercise, participants were divided into small groups, and each group was assigned one of IBM's quantum computers.  The teams constructed quantum circuits and submitted them as jobs using Quantum Composer. We provided a worksheet detailing research objectives along with a series of questions to spark critical thinking as they progressed through the exercise.  We also provided spreadsheets to record and evaluate their results. 
 


Specifically, the camp participants were asked to evaluate CNOT error and readout error for each experimental device to predict which computer would perform best. 
Then the participants proceeded to run 5 experiments for each single qubit and multi-qubit quantum circuit on their assigned quantum computer. The multi-qubit quantum circuit was designed to generate Bell states to illustrate correlations in measurement outcomes. They recorded measurement outcomes for $|0\rangle$ and $|1\rangle$ in the spreadsheet for statistical analysis. 
Correlation in measurement outcomes revealed which two-qubit experiments produced entangled qubits.  Groups were asked to compare their measured error rates to the other groups' and to IBM's specified errors.  


Working through the the final project introduced the students and teachers to numerous concepts in QIS experimentation, but it also taught us which concepts and activities were the most challenging for them. Students best grasped how to characterize quantum devices and how to get results by running Quantum Composer. Quantum circuit implementation and device characterization proved to be one of the least challenging aspects of the activity due to the substantial preparation from the week's curriculum. 

In fact, it was using a spreadsheet and the concept of correlation for data analysis that presented the largest challenge, as most students had never used a spreadsheet prior to the research exercise. Therefore, our lesson not only taught quantum information concepts, but it also taught the students practical skills needed for quantitative tasks. Overall, correlation was the most difficult concept for them to grasp as statistical analysis is absent in most high school classrooms. However, throughout this camp and final project, the students gained a better understanding of correlation and how it connects with measurement of entangled qubits. These observations, and those in the previous paragraph, were based on the interactions between the instructors and students rather than a quantitative evaluation.

The main outcome of the IBM Q research exercise was reinforcement of the scientific method - hypothesis, experiment, analysis, results-driven hypothesis justification - through critical thinking and hands-on experimentation. Using the IBM quantum experience and spreadsheets also provided practical skills that the students could use as they progress into the next academic stage of their lives either in STEM or other related fields. 

\subsection{Additional Sessions}
\label{sec:othersessions}

In addition to the main modules described above, a number of mini-talks were incorporated throughout the week to provide participants with exposure to career pathways, follow-on opportunities, resource sharing, lab tours, etc.  Here we provide details on a few of these additional sessions.

\textbf{Vectors and the Bloch Sphere:}  Building upon concepts introduced during the superposition and Hadamard gate module, the topics of vectors and the Bloch sphere were reintroduced utilizing an interwoven format \cite{Birnbaum2013}. First, students were guided through contrasting addition of “normal numbers” with addition of walking directions in a hands-on manner. This form of highly active learning has been found effective in promoting science efficacy and attitude in informal science outreach \cite{Todd2018}. Next, students were guided through an analysis of the state-space of a classical coin: a line of mixed states connecting two pure endpoints (heads and tails). This idea was then incorporated into the larger state-space of a quantum coin, where the previously-learned superposition states along the equator were contrasted with the maximally mixed state at the sphere’s center. Finally, students were introduced to vector manipulation on the sphere via rotations about an axis, connecting the visual representation to the fundamental Pauli single-qubit gates discussed previously during the module.

\textbf{Superconducting Qubits and Superconducting Testbed Virtual Lab Tour:}  After the participants had been introduced to IBM's Quantum Composer, this talk provided an ``under-the-hood" view of superconducting qubit operation and hardware.  We gave an overview of the basic principles and operation of superconducting qubits and used the virtual lab tour at DOE's Advanced Quantum Testbed \cite{AQT} to introduce participants to typical hardware and lab environments for this qubit platform.

\textbf{Atomic Qubits and Trapped Ion Virtual Lab Tour:}  To tie in the light lab activities performed earlier in the week, we also gave participants an introduction to atomic qubits, which included the basics of confining an ion with oscillating voltages, Doppler/laser cooling, and $\pi$/2 pulses to create superposition states.  We showed photographs of laboratory systems, single ions, and chains of ions.  With support from the UC Berkeley trapped ion group, we concluded this mini-talk with a live walk-through of their lab.

\textbf{Big Picture:} On the final day of the teacher and student camps, there was a ``big picture'' mini-talk that provided a high-level survey of the history, development, and current status of quantum computing. The aim was to put the topics that were discussed during the week within the context of the large, dynamic, world-wide QIST research effort and to connect the concrete progress made by the camp participants to the current state of the field, e.g., to show that some of the activities they engaged in during the week such as the IBM Q experiments are also part of current QIST research practice. This talk concluded by encouraging participants to find out more about QIST and join the research community in this exciting endeavor to build and leverage a quantum computer.

\textbf{Near-Peer Mentorship:} As a closing activity, one of the instructors (a junior post-doc) led a near-peer mentorship session \cite{Files2008}, answering student questions about education and careers in quantum information. As a transgender scientist, the instructor was also able to speak to the experience of working in QIST as multiple genders, as well as provide tools for resilience for all students. LGBTQ+ belonging in QIST was heavily emphasized. This session was especially important for transgender and gender-non-conforming students, who exhibit lower persistence in STEM than their cisgender peers \cite{Maloy2022}.

\subsection{Incorporating QIST in the High School Classroom}

To aid teachers in incorporating QIST lesson modules in their classrooms, each topic covered during QCaMP was linked with relevant Next Generation Science Standards (NGSS) \cite{NGSS}. NGSS are currently adopted by twenty states in the United States of America, in addition to twenty-four states that have science standards influenced by the NGSS framework. In particular, the focus was to expand on NGSS Disciplinary Core Idea (DCI) PS4: Waves and Their Applications in Technologies for Information Transfer. For high school students, this DCI includes standards that cover classical waves concepts such as mechanical waves and electromagnetic waves, as well as a conceptual introduction to the wave-particle duality. By expanding on this DCI, teachers can build upon classical physics already covered in their classrooms to transition into quantum concepts. In addition to NGSS, topics were also linked with Quantum Information Science (QIS) Key Concepts for K-12 Physics from the National Q-12 Education Partnership \cite{Q12Frame}, as well as Advanced Placement (AP) Physics 2 learning objectives from the College Board.  Table \ref{standards} clarifies which frameworks, standards, and learning objectives are linked to topics covered in QCaMP.

\begin{table*}[htbp]
\caption{QCaMP Topics Linked to Standards}
\begin{center}
\begin{tabular}{|c|c|c|c|}
\hline
\multirow{2}{*}{\textbf{Topic}} & \textbf{QIS Key concepts for K-12} & \textbf{Next Generation} & \textbf{AP Physics 2}\\
& \textbf{Physics Learner Outcomes} \cite{Q12Frame} & \textbf{Science Standards} \cite{NGSS} & \textbf{Learning Objectives} \cite{APPhys}\\
\hline
\multirow{4}{*} {Electromagnetic Waves} & \multirow{4}{*}{No reference as classical topics} & \multirow{4}{*} {HS-PS4-1} & Topic 6.1: 6.A.2.2;  \\
& & & Topic 6.2: 6.F.1.1, 6.F.2.1; \\
& & & Topic 6.3: 6.B.3.1; Topic 6.6: 6.C.1.1, \\
& & &  6.C.1.2, 6.C.2.1, 6.C.3.1, 6.C.4.1\\
\hline
\multirow{2}{*} {Wave-Particle Duality} & \multirow{2}{*} {2.1A,B,C; 2.2A} & \multirow{2}{*} {HS-PS4-3} & Topic 7.5: 1.D.1.1, 6.G.1.1, 6.G.2.1, \\
& & & 6.G.2.2; Topic 7.6: 6.F.4.1 \\
\hline
Polarization of Light, & \multirow{1.5}{*} {2.3A; 2.4A,C; 3.1A; 3.2A;} & \multirow{3}{*} {HS-PS4-3} & \multirow{3}{*} {Topic 7.7: 7.C.1.1} \\
Superposition of Quantum States, & \multirow{1.5}{*} {3.3A; 3.4A; 3.5A,B; 3.6C} & & \\
and Quantum Measurement & & & \\
\hline
Quantum Computing: Classical & 4.1A; 4.2A; 4.3B; & \multirow{2}{*} {HS-PS4-2, HS-PS2-6} & \multirow{2}{*} {Topic 7.5: 1.D.1.1, 6.G.1.1} \\
vs. Quantum Bits ("Qubits") & 4.4A,B; 7.1A,B,C,D,E & & \\
\hline
Entanglement & 5.1A, 5.2A, 5.2C, 5.3A, 5.4A & No reference & No reference \\
\hline
Quantum Computing: Logic and Gates & 7.1A,B; 7.3A,B,C,D,E,F; 7.4A,B,C & No reference & No reference \\
\hline
Quantum Computing: Applications & 7.5; 7.6A,B; 7.7; 7.8 & HS-PS4-2, HS-ETS1-4 & No reference \\
\hline
\end{tabular}
\label{standards}
\end{center}
\end{table*}

During the teacher camp, debrief sessions were intentionally placed throughout the week after each lesson module so teachers could have the opportunity to discuss with each other and the instructors their perspectives on the lessons. Debriefs were structured so that teachers could share key concepts they learned, how they would modify the lesson modules so they could meet the needs of their students, and any lingering questions. At the end of the week, teachers also had the opportunity to share additional resources and strategies with each other to support student success in the classroom. The debrief sessions not only enabled the teachers to reflect on their learning of the material, but also helped us identify adjustments for future iterations of the lessons. 

\subsection{Materials}
\label{sec:materials}

To facilitate hands-on learning despite the virtual format of the camp, we sent all participants a kit of materials prior to the camp start date.  Materials included necessary items for the Probability and Statistics module (Section \ref{sec:probability}), Light Labs modules (Section \ref{sec:lightlabs}), and a copy of the Q is for Quantum book \cite{Qi4Q} that informed the box gate formalism of the Classical Bits and Gates (Section \ref{sec:bits}) and Superposition and the Hadamard Gate (Section \ref{sec:superposition}) modules.  For convenience, we include a list of materials, vendors, and rough costing in Table \ref{materials} below.

\begin{table}[htbp]
\caption{QCaMP 2022 Materials}
\begin{center}
\begin{tabular}{|c|c|c|}
\hline
\textbf{Item Description} & \textbf{Cost} & \textbf{Vendor}\\
\hline
2x Dice & \$0.11 each & amazon.com\\
\hline
3x Polarizers & \$0.74 each & arborsci.com\\
\hline
1x Laser Diffraction Kit & \$118.65 each & carolina.com\\
\hline
4x AAA Batteries & \$1.03 each & staples.com\\
\hline
1x Q is for Quantum & \$12.99 each & amazon.com\\
\hline
\textbf{Total} & \textbf{\$138.20 per participant} & \\
\hline
\end{tabular}
\label{materials}
\end{center}
\end{table}

\section{Outcomes}
\label{sec:outcomes}

\subsection{Methods}

To evaluate impact on teacher and student attitudes about camp goals, the evaluation team worked with the camp team to design a survey with questions that reflected these areas. The survey was administered online at the end of each teacher and student camp through the Research Electronic Data Capture (REDCap) software \cite{Harris2009}. Evaluation was approved through the local Institutional Review Board (IRB). Questions asked about demographic areas and whether respondents felt they gained certain skills from the camp. Participants also had opportunities to provide open-ended feedback about their experiences.

\subsection{Diversity}

Evaluation captured feedback from a diverse set of participants within both teacher and student cohorts.  

\subsubsection{Teachers}

In total, 15 middle and high school teachers completed evaluation surveys. The primary subject areas of these teachers included science (e.g. biology, chemistry, computer science, physics) as well as engineering, math, and robotics. Their years of experience as a classroom teacher ranged from 1 to 34 years with the average being 12 years.  Demographic questions captured teachers’ race/ethnicity, as presented in Table \ref{T_RE} below. 7 teacher participants identified with underrepresented racial groups in STEM:  6 participants identified as Hispanic or Latino and 1 identified as Native Hawaiian or Other Pacific Islander. Regarding gender identity, 9 teachers identified as male and 6 identified as female. 

\begin{table}[h]
\caption{Teacher Survey Responses - Race/Ethnicity}
\begin{center}
\begin{tabular}{|c|c|}
\hline
\textbf{Race/Ethnicity} & \textbf{Count}\\
\hline
American Indian or Alaskan Native & 0\\
\hline
Asian & 4\\
\hline
Black or African American & 0\\
\hline
Hispanic or Latino & $\textbf{6}$ \\
\hline
Native Hawaiian or Other Pacific Islander & 1 \\
\hline
White & 4\\
\hline
$\textbf{Total}$ & $\textbf{15}$ \\
\hline
\end{tabular}
\label{T_RE}
\end{center}
\end{table}

\subsubsection{Students}

The 20 QCaMP student participants who completed an evaluation survey also represented diverse identities.  Table \ref{S_RE} below shows the students’ race/ethnicity responses.  7 student participants identified with underrepresented racial groups in STEM: 2 identified as American Indian or Alaska Native, 2 as Black or African American, and 3 as Hispanic or Latino. Regarding gender identity, 11 students identified as male, 8 as female, and 1 as gender non-binary. 

\begin{table}[h]
\caption{Student Survey Responses - Race/Ethnicity}
\begin{center}
\begin{tabular}{|c|c|}
\hline
\textbf{Race/Ethnicity} & \textbf{Count}\\
\hline
American Indian or Alaska Native & 2\\
\hline
Asian & $\textbf{10}$\\
\hline
Black or African American & 2\\
\hline
Hispanic or Latino & 3 \\
\hline
Native Hawaiian or Other Pacific Islander & 0 \\
\hline
White & 1\\
\hline
Multi-Race & 2\\
\hline
$\textbf{Total}$ & $\textbf{20}$ \\
\hline
\end{tabular}
\label{S_RE}
\end{center}
\end{table}

As the camp was open to high school students in the summer, students were asked what grade level they would enter in the following fall.  Nearly half of the students ($n = 9$) reported entering the 12th grade; the responses are provided in Table \ref{S_GL} below. 

\begin{table}[h]
\caption{Student Survey Responses - Grade Level in Fall 2022}
\begin{center}
\begin{tabular}{|c|c|}
\hline
\textbf{Grade Level} & \textbf{Count}\\
\hline
10th & 3\\
\hline
11th & 8\\
\hline
12th & \textbf{9}\\
\hline
$\textbf{Total}$ & $\textbf{20}$ \\
\hline
\end{tabular}
\label{S_GL}
\end{center}
\end{table}

\begin{table}[h]
\caption{Student Survey Responses - Parents/Guardians Education}
\begin{center}
\begin{tabular}{|c|c|}
\hline
\textbf{Parents/Guardians Earned College Degree} & \textbf{Count}\\
\hline
Yes & \textbf{12}\\
\hline
No & 7\\
\hline
Unsure & 1\\
\hline
$\textbf{Total}$ & $\textbf{20}$ \\
\hline
\end{tabular}
\label{S_PGE}
\end{center}
\end{table}

To capture college generational status, students were asked if either of the student’s parent(s) or guardian(s) earned a college degree. The responses are tallied in Table \ref{S_PGE}, with the majority of students ($n = 12$) indicating that “yes,” one of their parent(s) or guardian(s) earned a college degree.  7 students responded “no,” which means that if they choose to pursue a post-secondary education, they would be a first-generation college student. 

For the questions in Tables \ref{T_IQ} and \ref{S_IQ}, there were five levels of agreement for the response options that ranged from Strongly Disagree to Strongly Agree with a Neutral option in the middle. To calculate averages, all responses were given a value 
ranging from 1 (assigned to responses of ``Strongly Disagree") to 5 (assigned to responses of ``Strongly Agree").
The questions were prefaced with “As a result of my attendance at QCaMP…” to help participants focus on impact from the camp experience specifically.

\subsection{Teacher Camp Evaluation}

Overall, teachers reported strong positive impacts of the camp. In particular, they felt more confident delivering instruction in areas such as quantum and physics, that their students would benefit from teachers attending QCaMP, and that they learned new ways to engage students in QCaMP material. All questions and averages can be found in Table \ref{T_IQ}.

\begin{table*}[htbp]
\caption{Averages for Teacher Survey Responses - Impact Questions}
\begin{center}
\begin{tabular}{|c|c|}
\hline
\textbf{Teachers} & \textbf{Avg}\\
\hline
I feel more confident in delivering instruction to my students in quantum technologies. & 4.07\\
\hline
I feel more confident in delivering instruction to my students in computing. & 4.07\\
\hline
I feel more confident in delivering instruction to my students in mathematics. & 4.00\\
\hline
I feel more confident in delivering instruction to my students in physics. & 4.07\\
\hline
I am likely to implement QCaMP material in my classroom this coming academic year. & 4.00\\
\hline
I am likely to share QCaMP material with my colleagues this coming academic year. & 4.00\\
\hline
I learned new ways to engage my students in QCaMP material. & 4.07\\
\hline
My interest in STEM has increased. & 4.07\\
\hline
My students will benefit from my QCaMP experience. & 4.13\\
\hline
I am interested in participating in additional QCaMP programs. & 4.00\\
\hline
I am interested in continued engagement and education in QIST. & 3.93\\
\hline
\end{tabular}
\label{T_IQ}
\end{center}
\end{table*}

There were also opportunities for teachers to answer open-ended questions. For example, they were asked why they decided to pursue QCaMP, and most teachers reported wanting to expand their knowledge regarding quantum computing and to be able to implement it in classrooms or extracurricular activities. The following are some of the teachers’ direct responses: 
\begin{itemize}
    \item \textit{``I'm going into my first year of teaching and I thought it wold [sic] be extremely helpful to be informed about the research being done currently and how I could create lessons that inform students about it!"}
    \item \textit{``Several students in my STEM club have wanted to learn more about quantum computing, and I've always been curious about how to code for them."}
\end{itemize}

Regarding what elements of the camp were enjoyed the most, teachers appreciated the subject experts (scientists) involved in the camp and the interaction with other teachers. Finally, there was an opportunity to provide any other feedback; the following are some of the responses:   

\begin{itemize}
    \item \textit{``An overall excellent experience, thank you for the opportunity to uncover so much I knew nothing about and get me excited about the future of computing!"}
    \item \textit{``This was an amazing camp! Thank you so much for informing me about all these topics and connecting me with other teachers across the US! It was so much fun and I know the kids will equally enjoy this camp!"}
\end{itemize}

\subsection{Student Camp Evaluation}

Students also reported an overall positive experience, particularly when asked if they understand more about quantum computing, how mistakes form part of the scientific process, and if they are more likely to take a class in related subjects during the upcoming school year.  All questions and averages can be found in Table \ref{S_IQ}.

\begin{table*}[htbp]
\caption{Averages for Student Survey Responses - Impact Questions}
\begin{center}
\begin{tabular}{|c|c|}
\hline
\textbf{Students} & \textbf{Avg}\\
\hline
I understand more about quantum science and technologies than I did before the camp. & 4.55\\
\hline
I understand more about computing than I did before the camp. & 4.30\\
\hline
I understand more about mathematics than I did before the camp. & 3.32\\
\hline
I understand more about physics than I did before the camp. & 4.05\\
\hline
I am more comfortable asking questions about material than I was before the camp. & 3.25\\
\hline
I am more comfortable making mistakes on tasks than I was before the camp. & 3.15\\
\hline
I understand that failing and mistakes are a part of the scientific process. & 4.45\\
\hline
I am more likely to take a class in quantum, computing, mathematics, and/or physics this upcoming school year. & 4.25\\
\hline
I feel more comfortable working on a team with my peers. & 3.50\\
\hline
I understand how I contribute to teams that I am a part of. & 4.05\\
\hline
I learned new team work skills I can use in school. & 3.60\\
\hline
When I get back to school this year, I am going to tell all my friends about QCaMP. & 3.95\\
\hline
My interest in pursuing a STEM career has increased. & 4.05\\
\hline
My interest in pursuing a quantum science related career has increased. & 3.45\\
\hline
I am interested in participating in additional QCaMP programs. & 4.10\\
\hline
I want to be a student leader at next year's QCaMP. & 3.25\\
\hline
I feel a sense of belonging to people with similar interests to me. \textit{*1 missing response} & 4.05\\
\hline
\end{tabular}
\label{S_IQ}
\end{center}
\end{table*}

Similar to teacher evaluations, students had the opportunity to answer a number of open-ended questions. They reported a variety of aspects they enjoyed most out of the camp, including meeting the subject experts (scientists), participating in experiments, and learning about career opportunities. The following are some of their responses: 

\begin{itemize}
    \item \textit{``I got to learn a lot about things I don't learn about in school, and it was really cool."}
    \item \textit{``Getting to see real scientists explain quantum mechanics and computers to us."}
\end{itemize}

For any additional feedback, students provided suggestions for future QCaMPs, particularly suggesting that we host an in-person camp. Other responses reflected the positive time the students seemed to have. The following are some of their responses: 

\begin{itemize}
    \item \textit{``Overall it is a good experience even though it was defienetly [sic] new for me."}
    \item \textit{``It was fun and I really learned a lot"}
\end{itemize}

\section{Outlook}
\label{sec:outlook}

Due to the overwhelmingly positive evaluation feedback and continued growth in the QIST field, we will build upon 2022's developed curriculum to offer QCaMP 2023.  This next camp has a few notable differences from 2022.

Firstly, by popular request, in addition to a virtual option, we will also host in-person cohorts in regions with multiple participants.  There will be an in-person cohort of teachers in Albuquerque, NM, and in-person cohorts of students in Albuquerque and Santa Fe, NM. The virtual and in-person camps will take place simultaneously -- one week for teachers and another week for students -- with virtual instruction by QIST researchers to limit demand on researcher time.  We are anticipating 24 teachers and 50 students in 2023.

Secondly, we are adjusting the 2023 curriculum to incorporate even more time and activities on IBM's Quantum Composer as it was one of the more popular activities from 2022.  In addition to the research project, hands-on IBM Q exercises will be included in the Probability and Statistics module and the Superposition module. This change will not only help reinforce concepts through active learning, but also underscore connections between topics across the curriculum. To accommodate, the Intro to IBM Q module will occur on Monday afternoon, swapping with the Probability and Statistics module. We also decided to spread the Light Labs over two days to better balance experimental activities and screen time. An overview of the 2023 curriculum can be found in Table \ref{2023Curr}.

Thirdly, we are excited to partner with SparCQS, an NSF Center for Quantum Networks initiative, led and operated out of Northern Arizona University.  Expanding the reach of a traditional outreach program, SparCQS includes a mobile quantum laboratory bringing ‘hands-on’ quantum science experiences directly to schools and communities \cite{sparcqs}. SparCQS will join our in-person Albuquerque cohorts on Friday of both the teacher and student camps.  During this time, we will introduce our virtual participants to various online quantum games.

\begin{table*}[htbp]
\caption{QCaMP 2023 Curriculum}
\begin{center}
\begin{tabular}{|c|c|c|c|c|c|}
\hline
& \textbf{Monday} & \textbf{Tuesday} & \textbf{Wednesday} & \textbf{Thursday} & \textbf{Friday}\\
\hline
\multirow{2}{*}{\textbf{AM}} & \multirow{2}{*}{Classical Bits and Gates} & \multirow{2}{*}{Polarization} & \multirow{2}{*}{Double Slit Experiment} & Entanglement \& & \multirow{2}{*}{IBM Q Research Project} \\
& & & & IBM Q Research Project &\\
\hline
\multirow{2}{*}{\textbf{PM}} & \multirow{2}{*}{Intro to IBM Q} & Superposition & Probability and Statistics &  Virtual Lab Tours \& & SparCQS \\
& & \& IBM Q & \& IBM Q & IBM Q Research Project & \& Quantum Games\\
\hline
\end{tabular}
\label{2023Curr}
\end{center}
\end{table*}

We would like to expand further in future years by offering in-person cohorts across the country as well as virtual options to continue to make QCaMP accessible to as many underrepresented communities as possible.  To be successful, we will need the help of our broader QIST community to recruit students in their regions, to identify classroom spaces for hosting in-person cohorts, to facilitate in-person cohorts (student supervision, tech support, etc.), and to help identify potential financial sponsors for their region's participant stipends.  As another option, we are moving toward making all of our material (lesson slides, worksheets, and video recordings) freely available for anyone who wishes to use them in their own classrooms or camps.

Two of the most impactful outcomes of this effort have been sparking further curiosity by students into more complicated concepts in quantum information science and creating a channel between expert QIST researchers and prospective QIST students. Following a pilot program at JSTI Virtual 2021 \cite{jsti}, wherein we used a similar curriculum spread over a two-week virtual program, one of the students worked with an organizer on science fair projects investigating quantum teleportation and quantum machine learning. One of these projects received numerous regional, national, and international awards, opening the previously unattainable opportunity to gain acceptance to top US undergraduate institutions. This student will be starting in a top undergraduate physics program focusing on QIST. This example demonstrates that it is possible to teach quantum computing to high school age students, and the outcomes can be profound.

\section*{Acknowledgment}


We acknowledge leadership activities performed by Amy Tapia of the Community Engagement Office at Sandia National Laboratories, Faith Dukes of the K-12 Programs at Lawrence Berkeley National Laboratory, and Yolanda Lozano of the Computer Science Alliance; and classroom facilitation by Emily Clauss of La Cueva High School and Cari Hushman of the Department of Education at the University of New Mexico.  Additional QCaMP mini-talk speakers were Will Kindel (IQM), Francis Vigil (NIEA), Kayla Lee (IBM), and Lauren Thomas Quigley (IBM).  Thanks to the Haeffner Group at UC Berkeley for facilitating a virtual trapped ion lab tour.

This material is based upon work supported by the U.S. Department of Energy, Office of Science, National Quantum Information Science Research Centers, Quantum Systems Accelerator. Additional support is acknowledged from Sandia National Laboratories, Lawrence Berkeley National Lab, and IEEE.  Tz. B. Propp acknowledges funding from National Science Foundation Grant No. PHY-1630114.

This article has been co-authored by employees of National Technology \& Engineering Solutions of Sandia, LLC under Contract No. DE-NA0003525 with the U.S. Department of Energy (DOE). The authors own all right, title and interest in and to the article and are solely responsible for its contents. The United States Government retains and the publisher, by accepting the article for publication, acknowledges that the United States Government retains a non-exclusive, paid-up, irrevocable, world-wide license to publish or reproduce the published form of this article or allow others to do so, for United States Government purposes. The DOE will provide public access to these results of federally sponsored research in accordance with the DOE Public Access Plan \url{https://www.energy.gov/downloads/doe-public-access-plan}.

This paper was prepared for information purposes, and is not a product of HSBC Europe or its affiliates. Neither HSBC Europe nor any of its affiliates make any explicit or implied representation or warranty and none of them accept any liability in connection with this paper, including, but not limited to, the completeness, accuracy, reliability of information contained herein and the potential legal, compliance, tax or accounting effects thereof.
\bibliographystyle{unsrt}
\bibliography{references}

\vspace{12pt}

\end{document}